\documentstyle[psfig,floats,aps,aipbook]{revtex}

\newcommand{\figside}[3]{
\begin{figure}[t]
\begin{center}
\begin{tabular}{lp{0.6\textwidth}}
\raisebox{-2.0in}{#1} &
{#2}
\end{tabular}
\end{center}
\refstepcounter{figure} #3
\end{figure}
}

\def\kms{\hbox{km~s$^{-1}$}}
\def\mathnew{\mathsurround=0pt}
\def\simov#1#2{\lower .5pt\vbox{\baselineskip0pt \lineskip-.5pt
        \ialign{$\mathnew#1\hfil##\hfil$\crcr#2\crcr\sim\crcr}}}
\def\simgreat{\mathrel{\mathpalette\simov >}}
\def\simless{\mathrel{\mathpalette\simov <}}

\begin{document}
\title{Constraints on the Galactic Corona Models of Gamma-Ray Bursts
From the 3B Catalogue}
\author{Tomasz Bulik$^{1,2}$ and Donald Q. Lamb$^1$}
\address{$^1$ Department of Astronomy and Astrophysics\\
University of Chicago\\
5640 South Ellis Avenue,
Chicago, IL 60637}
\address{$^2$ Nicolaus Copernicus Astronomical Center\\
           Bartycka 18, 00-716 Warsaw, Poland}
\maketitle
\begin{abstract}
We investigate the viability of Galactic corona models of gamma-ray
bursts by calculating the spatial distribution expected for a
population of high-velocity neutron stars born in the Galactic disk and
moving in a gravitational potential that includes the Galactic bulge,
disk, and a dark matter halo.  We consider models in which the bursts
radiate isotropically and in which the radiation is beamed.  We place
constraints on the models by comparing the resulting brightness and
angular distributions with the data in the BATSE 3B catalog.  We find
that, if the burst sources radiate isotropically, the Galactic corona
model can reproduce the BATSE peak flux and angular distributions for
neutron star kick velocities $\simgreat$ 800 km s$^{-1}$, source
turn-on ages $\simgreat$ 20 Myrs, and BATSE sampling distances 130~kpc
$\simless d_{\rm max} \simless$ 350~kpc.  If the radiation is beamed,
no turn-on age is required and agreement with the BATSE data can be
found provided that the width of the beam is $\simless 20^\circ$.
\end{abstract}
%


\begin{figure}[t]
\centerline{~~~~}
\vspace{-10mm}
\begin{center}
\begin{tabular}{lr}
a){\psfig{file=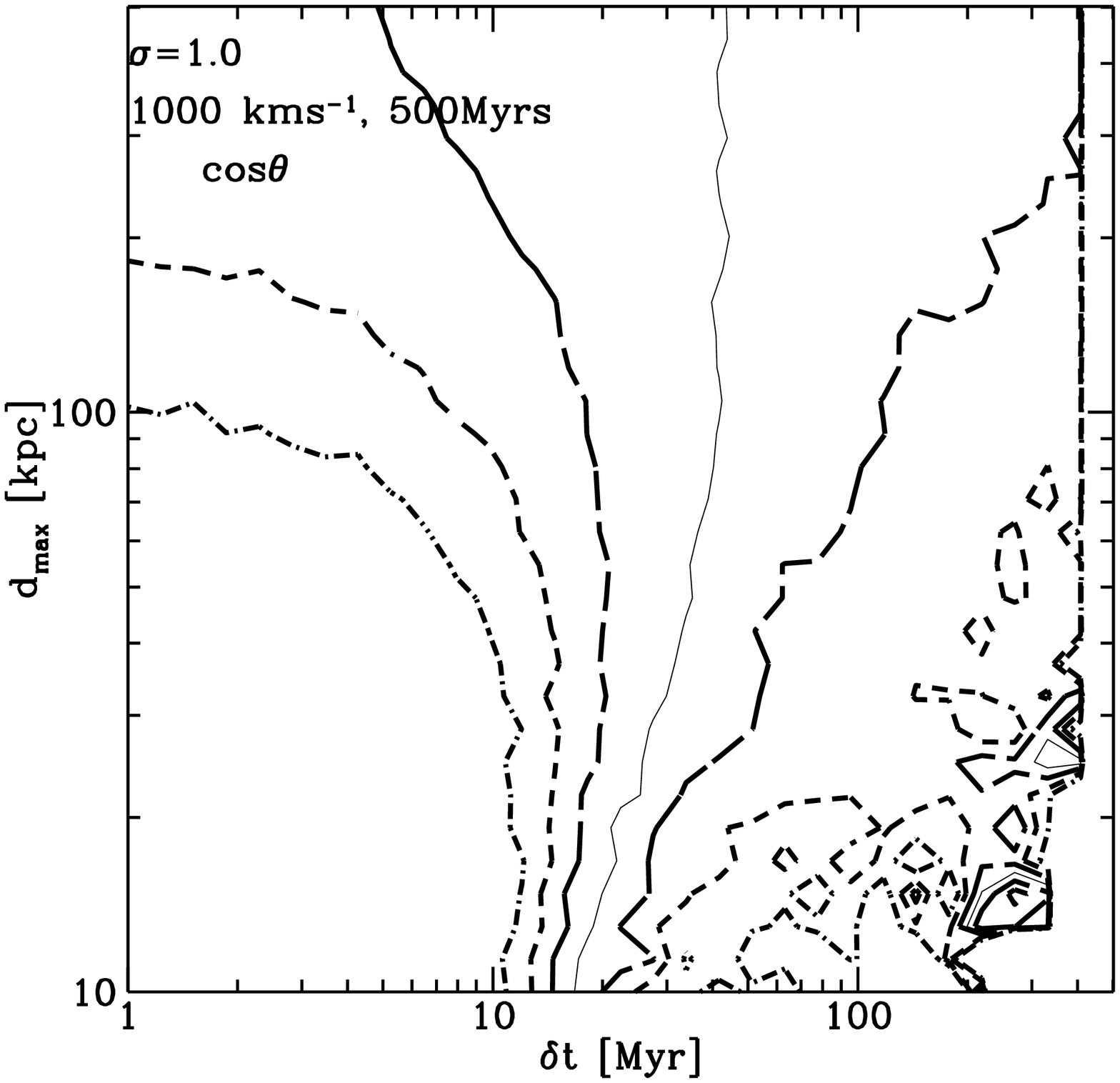,width=5.4cm}} &
b){\psfig{file=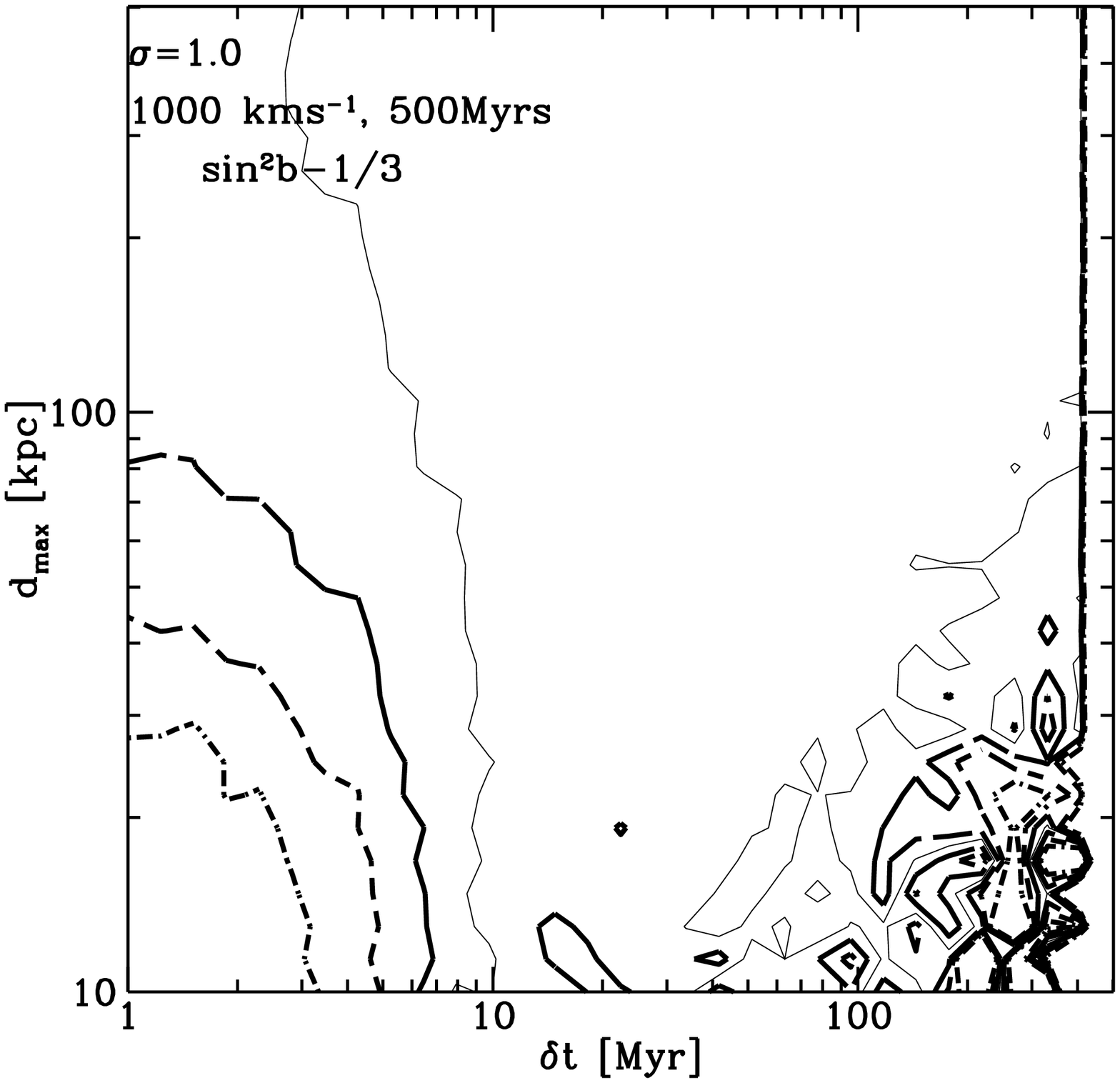,width=5.4cm}} \\
c){\psfig{file=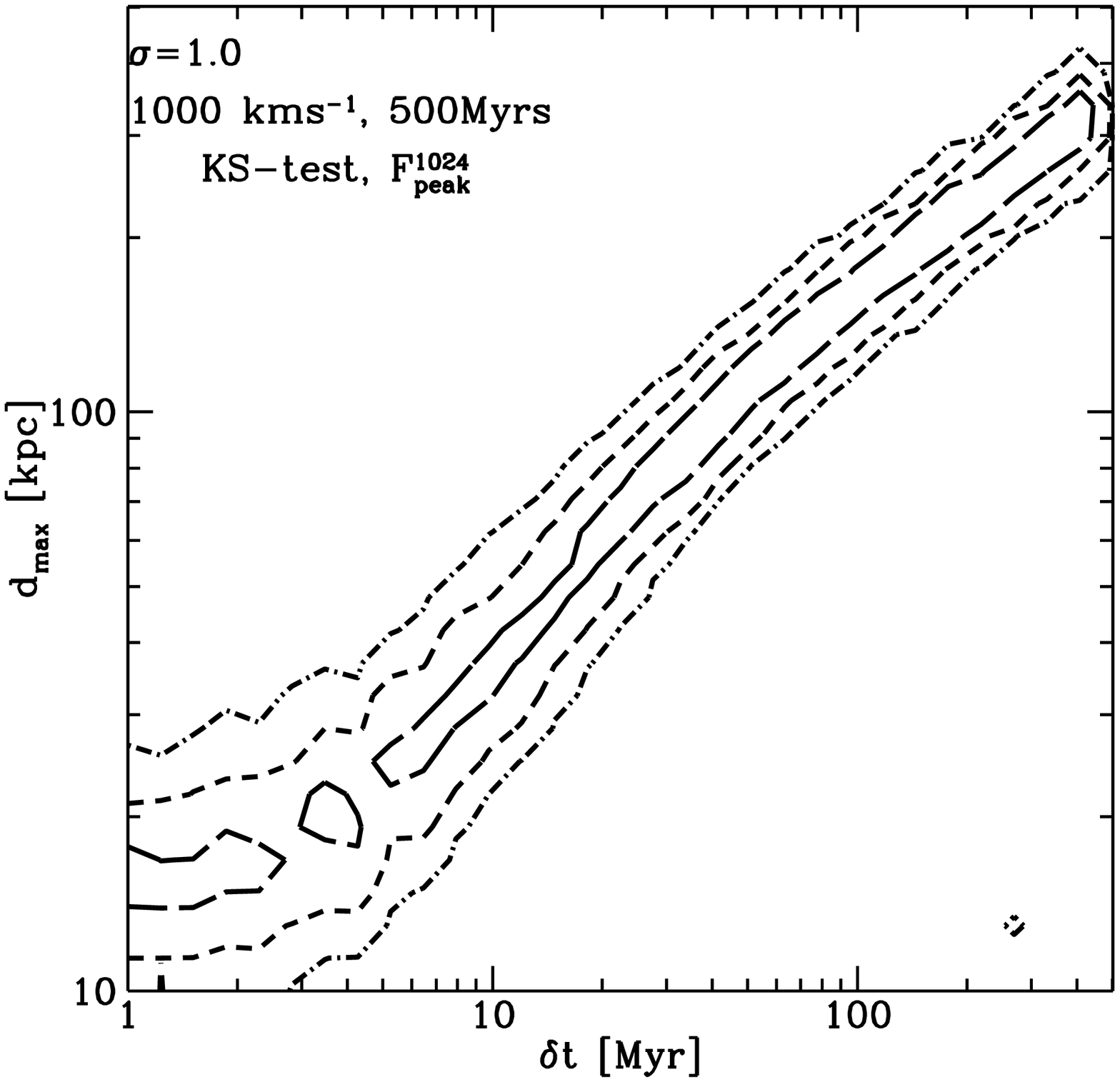,width=5.4cm}}  &
d){\psfig{file=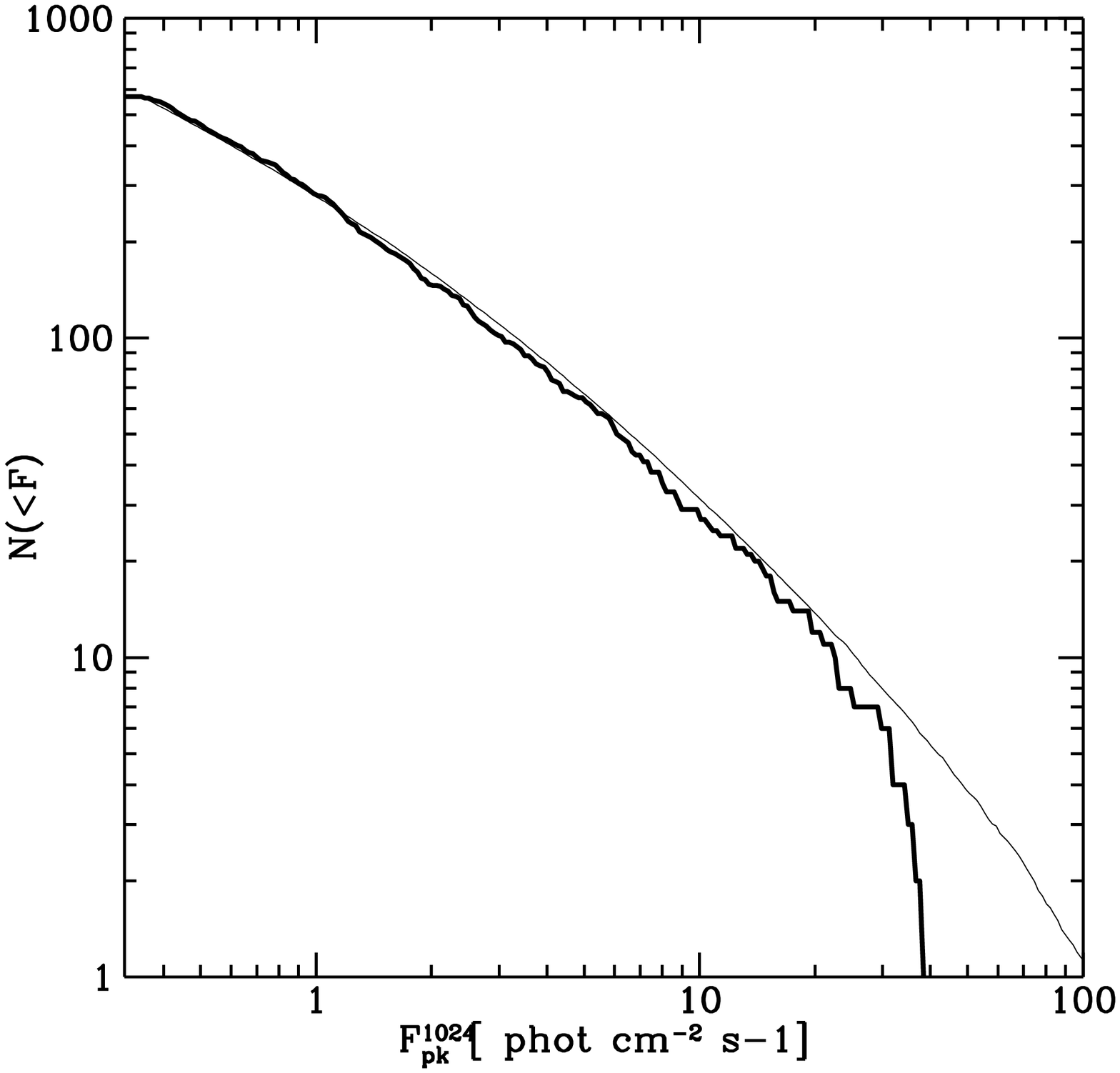,width=5.4cm}} 
 \end{tabular}
\end{center}
\caption{Comparison of a Galactic halo model in which neutron stars
are born with a kick velocity of
$1000$~\kms, have a burst-active phase lasting
$\Delta t = 5\times 10^8$ years, and a luminosity function with
width $\sigma=1.0$, with a self-consistent sample of 570
bursts from the BATSE~3B catalogue.  Panels (a) and (b) show the
contours in the ($\delta t$, $d_{\rm max}$)-plane along which the
Galactic dipole and quadrupole moments of the model differ from those
of the data by $\pm$ 1$\sigma$ (solid lines), $\pm$ 2$\sigma$ (dashed
line), and $\pm$ 3$\sigma$ (short-dashed line) where $\sigma$ is the
model variance; the thin line in panels (a) and (b) show the contour where the
dipole and quadrupole moments for the model equals that for the data.  Panel (c) shows
the contours in the ($\delta t$, $d_{max}$)-plane along which 32\%,
5\%, and $0.4\% $ of simulations of the cumulative
distribution of 570 bursts drawn from the peak flux distribution of the
model have KS deviations $D$ larger than that of the data. Panel (d)
shows brightness distribution of a model with $\delta t=30$~Myrs and 
$d_{\rm max}=80$~kpc and the BATSE  data. 
}
\vspace{-5mm}
\end{figure}

\section{Introduction}

Gamma-ray bursts (GRB's) continue to confound astrophysicists nearly a
quarter century after their discovery \cite{Klebesadel73}.
Before the launch of CGRO, most scientists thought that GRB's came from
magnetic neutron stars residing in a thick disk (having a scale height
of up to $\sim$ 2 kpc) in the Milky Way  \cite{Higdon90}.
The data gathered by BATSE showed the existence of a
rollover in the cumulative brightness distribution of GRB's and
that the sky distribution of even faint GRB's is consistent with
isotropy \cite{Briggs95}.

 Galactic models attribute the bursts primarily to
high-velocity neutron stars in an extended Galactic halo, which must
reach one fourth or more of the distance to M31 ($d_{\rm M31} \sim
690$ kpc) in order to avoid any discernible anisotropy  
\cite{Hakkila94,Hartmann94,Bulik95}.
 Cosmological models place
the GRB sources at distances $d\sim 1-3$~Gpc, corresponding to
redshifts $z\sim 0.3 - 1$;  a source population at such large distances
naturally produces an isotropic distribution of bursts on the sky.  In
addition, studies show that the expansion of the universe can reproduce
the observed rollover in the cumulative brightness distribution  
e.g., \cite{Fenimore93}.

Recent studies \cite{Lyne94,Frail94}
have revolutionized our understanding of the birth velocities of radio
pulsars.  They show that a substantial fraction of neutron stars have
velocities that are high enough to produce an extended halo around the
Milky Way like that required by Galactic halo models of GRBs  \cite{Li92}.  

Detailed studies of the spatial distribution expected for high-velocity
neutron stars born in the Galactic disk \cite{Bulik95,Podsiadlowski95}
show that there is a large region in the parameter space
where galactic models are consistent with the data from the 2B catalogue.
The aim of this work is to re-evaluate this models in the light of the
BATSE~3B catalogue.

\section{Models}

  We have calculated detailed models of the spatial
distribution expected for a population of high-velocity neutron stars
born in the Galactic disk and moving in a Galactic potential that
includes the bulge, disk, and a dark matter halo.  

We use the mass distribution and potential~\cite{Kuijken89} 
which includes a doubly exponential disk,
a bulge, and a dark matter halo \cite{Bulik95}.
 The circular velocity $v_c$ and the Galactic disk lead to
characteristic angular anisotropies as a function of burst brightness
which provide a signature, and therefore a test, of high-velocity
neutron star models. 

We assume that the neutron stars are born with the local 
circular velocity
$v_c \approx 220~\kms$ of the Galactic disk and an isotropic
distribution of initial kick velocities $v_{\rm kick}$ ranging from
$200$ to $1200$~\kms.  We follow the resulting orbits for up to $3
\times 10^9$ years.  Given that current knowledge of $v_{\rm kick}$ is
poor, we adopt a Green's-function approach: we calculate the
spatial distribution of neutron stars for a set of kick velocities
(e.g., $v_{\rm kick} = 200, 400,..., 1200$~\kms).

We consider a model in which the bursts are standard
candles, i.e. $L = \delta(L-L_0)$, and also models 
with the log-normal luminosity function with some width,
$P(L,L+dL)\approx \exp(-(\log(L/L_{av})/\sigma)^2)dL/L$. 
We denote $d_{av}$ the distance to an average burst. If the width
of the luminosity function is small than $d_{av}$ tends to 
$d_{max}$ - the sampling distance. 
We parametrize the burst-active
phase by a turn-on age $\delta t$ and a turn-off time $\Delta t$, and assume
that the rate of bursting is constant throughout the burst-active
phase; i.e., the burst rate $r=$~const for $\delta t \le t \le
\Delta t$ and $0$ otherwise.  The high-velocity neutron star model then
has the following parameters: $v_{\rm kick}$, $\delta t$, $\Delta t$, 
BATSE sampling depth $d_{\rm max}$, and the width of the luminosity function
$\sigma$. We also consider a beaming mode \cite{Li92} in which bursting
occurs in a cone with width $\theta_b$ along the initial kick velocity of a neutron star. Such models do not require a turn-on delay .

\section{Comparison between models and data}
We compare the models with a carefully-selected data set that is
self-consistent \cite{Bulik95}.
  We use only bursts that 
  a) trigger on the 1024~ms timescale, and have $t_{90} > 1024$~ms,
 b) have $F_{\rm pk}^{1024}$, the peak flux in 1024~ms, since we adopt
 it as the brightness measure.
 c) have $F_{\rm pk}^{1024} \ge 0.35$~photons~cm$^{-2}$~s$^{-1}$ in order
to avoid threshold effects \cite{Fenimore93,Zand94}.
  We also exclude overwriting bursts and MAXBC bursts. 
The 3B catalogue 
contains 570
bursts satisfying the above criteria; this set of bursts has Galactic
dipole and quadrupole moments $\langle \cos \theta \rangle =0.018 \pm
0.0241$, and $\langle \sin^2 b-{1\over 3} \rangle = -0.011 \pm 0.012$, and
$\langle \cos \theta_{M31} \rangle = 0.0078 \pm 0.0241$.

We test
the viability of Galactic halo models  by  comparing the Galactic
dipole and quadrupole moments, $\langle \cos \theta \rangle$ and
$\langle \sin^2 b - 1/3 \rangle$, of the angular distribution of bursts
for the model with those for the above set of bursts, using $\chi^2$.
We have also compared the peak flux distribution for the model with
that for the above set of bursts, using the KS test.

\begin{figure}[t]
\vspace{-5mm}
\begin{center}
\begin{tabular}{lcr}
\psfig{file=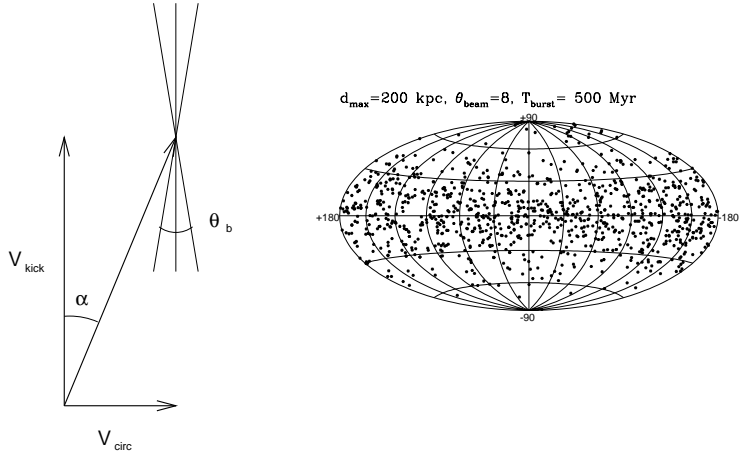,height=6cm,angle=270} & ~~~~ &
\psfig{file=bnlmap200.500t8,width=6cm}
\end{tabular}
\caption{The left panel shows the lower limit  on the beaming
angle $\theta_b$. A star with the kick velocity $V_{kick}$
perpendicular to the disk plane also has a galactic circular velocity
$V_{circ}$ in the plane, however, it is bursting along the 
direction of $V_{kick}$. If the angle $\alpha$ is larger than $\theta_b$
the bursts will never be seen by an observer in the Galaxy, relatively
close to the birthplace of the neutron star. The right panel shows a 
skymap for $v_{kick}=10^3$ kms$^{-1}$, and $\theta_b=8^\circ$.
 }
\end{center}
\vspace{-5mm}
\end{figure}

These comparisons do {\it not} provide estimates of model parameters
(i.e., they do not yield parameter confidence regions), but are meant
only to be a rough ``goodness-of-fit" guide to models which should be
tested using a more rigorous approach like the maximum likelihood
method.

In Figure~1 we present the results for a
Galactic halo model in which neutron stars are born with a uniform
single velocity  $1000$~\kms\, turn-off time $\Delta t = 5\times 10^8$~yrs,
and a log-normal luminosity function with $\sigma=1.0$.  As an example,
in the  model with $\delta t =100{~\rm Myrs}$ and $d_{\rm ave}=170{~\rm kpc}$ 
the expected dipole and quadrupole moments are
$\langle \cos\theta\rangle =0.0033$ and $\langle \sin^2 b-1/3\rangle = 
-0.0046$, after correcting for the BATSE exposure. This values are extremely
close to those expected for isotropy. 

The beaming model~\cite{Li92} fits the data when the beaming angle
$\theta_b\approx 20^\circ$, and the constraints on the BATSE sampling
distance are similar to the case of isotropic emission. An interesting feature
of the beaming model is that $\theta_b$ is bounded on both sides.
The upper limit is due to the galactic anisotropy  seen for young
neutron stars because of their birth places. The lower limit
is due to the galactic rotational velocity which leads to lack of 
bursts in galactic polar directions when the beaming angle is too small.
The lower limit is approximately $\tan\theta_b > v_c /v_{kick}$, 
see Figure~2.

Comparisons of this kind show that the high-velocity neutron star model
can reproduce the peak flux and angular distributions of the bursts in
the BATSE 3B catalogue for neutron star kick velocities $v_{\rm kick}
\simgreat 800$~\kms, burst turn-on ages $\delta t \simgreat
20$~million years, and BATSE sampling depths $130$~kpc $\simless d_{\rm
max} \simless 350$~kpc. It is clear from this 
comparison that global isotropy comparisons will not yield  the answer
to the question of the origin of gamma-ray bursts. 

In high-velocity neutron star models, the slope of the cumulative peak
flux distribution for the brightest BATSE bursts and the PVO bursts
reflects the space density of the relatively small fraction of burst
sources in the solar neighborhood.  The standard candle models reproduce
the BATSE peak flux distribution but they are unable to explain the 
$-{3\over 2}$ slope observed for the PVO bursts. The luminosity function
"fills" the void near us with distant bright bursts which appear as
isotropic, nearby bursts. Moreover, increasing the width  of the
luminosity function allows to relax the constraint on the
burst turn-on time $\delta t$.

M31 provides a strong constraint on the BATSE sampling distance $d_{\rm
max}$  \cite{Hakkila94}. 
 We have investigated the
effects of M31 within the framework of the high-velocity neutron star
model described above by including the distortion of the Galactic halo
potential due to M31, as well as the spatial distribution of burst
sources emanating from M31.  The results of this work
are summarized in these proceedings \cite{Coppi_here}.


\begin{references}

\bibitem {Klebesadel73}
Klebesadel, R. W., Strong, I. B., and Olson, R. A.  1973, {\  Ap.J.},
182, L85.

\bibitem {Higdon90}
Higdon, J. C., and Lingenfelter, R. E. 1990, {\  A.R.A.A.}, 28, 401.


\bibitem {Briggs95}
Briggs, M., et al., 1995, Ap.J, in press.

\bibitem {Hakkila94}
Hakkila,~J. et~al. 1994, {\  Ap.J.}, 422, 659.

 \bibitem {Hartmann94}
Hartmann, D., et al. 1994, {\  Ap.J. Suppl.}, 90, 893.

\bibitem {Bulik95}
Bulik, T. and Lamb, D.Q., 1995, A\& Sp.Sc., in press.

\bibitem {Fenimore93}
Fenimore, E. E., et al. 1993, Nature, 366, 40.


\bibitem {Lyne94}
Lyne,~A. G. and Lorimer,~D. R. 1994, {\  Nature}, 369, 127.


\bibitem {Frail94}
Frail,~D. A., Goss,~W. M., and Whiteoak,~J. B. 1994, {\  Ap.J.}, 437,
781.


\bibitem {Li92}
Li,~H., and Dermer,~C. 1992, Nature, 359, 514.

\bibitem {Podsiadlowski95}
Podsiadlowski,~Ph., Rees,~M. J., and Ruderman,~M. 1995, {\ 
M.N.R.A.S.}, 237,755.

\bibitem {Kuijken89}
Kuijken, K., and Gilmore, G. 1989, {\  M.N.R.A.S.}, 239, 571.

\bibitem {Zand94}
in~'t~Zand,~J. J. M. and Fenimore,~E. 1994, in Gamma-Ray Bursts,
AIP Conference Proceedings No.~307, ed.
ed.~G.~J.~Fishman, J.~J.~Brainerd, and K.~Hurley (New York: AIP), p. 692.

\bibitem{Coppi_here}
Coppi, P.S, Bulik, T., and Lamb, D.Q., these proceedings.


\end{references}
\end{document}